\documentclass[prl,twocolumn]{revtex4}
\usepackage{graphicx}
\usepackage{bm}
\begin{document}
\newcommand{\Sv}{{\bm S}}
\newcommand{\Tv}{{\bm T}}
\newcommand{\nv}{{\bm n}}
\newcommand{\Ne}{{N_{\rm e}}}
\newcommand{\eF}{{\epsilon_F}}
\newcommand{\Ev}{{\bm E}}
\newcommand{\kF}{{k_F}}
\newcommand{\kp}{{k_\perp}}
\newcommand{\Kp}{{K_\perp}}
\newcommand{\sigmav}{{\bm \sigma}}
\newcommand{\phiz}{{\phi_0}}
\newcommand{\rhow}{{\rho_{\rm w}}}
\newcommand{\Vz}{{V_0}}
\newcommand{\Mv}{{\bm M}}
\newcommand{\tiln}{{\tilde{n}}}
\newcommand{\tilv}{{\tilde{v}}}
\newcommand{\tilK}{{\tilde{K}_\perp}}
\newcommand{\Area}{A}
\newcommand{\dx}{{d^3 x}}
\newcommand{\xv}{{\bm x}}
\newcommand{\rv}{{\bm r}}
\newcommand{\kv}{{\bm k}}
\newcommand{\qv}{{\bm q}}
\newcommand{\Vv}{{\bm V}}
\newcommand{\vv}{{\bm v}}
\newcommand{\Av}{{\bm A}}
\newcommand{\Bv}{{\bm B}}
\newcommand{\av}{{\bm a}}
\newcommand{\DOS}{{\nu}}
\newcommand{\kB}{{k_B}}
\newcommand{\js}{{j_{\rm s}}}
\newcommand{\jsc}{{j_{\rm s}^{\rm cr}}}
\newcommand{\jsca}{{j_{\rm s}^{\rm cr (1)}}}
\newcommand{\jscb}{{j_{\rm s}^{\rm cr (2)}}}
\newcommand{\jc}{{j^{\rm cr}}}

\title{ Theory of Current-Driven Domain Wall Motion: \\ 
A Poorman's Approach} 
\author{Gen Tatara}
\affiliation{
Graduate School of Science, Osaka University, Toyonaka, Osaka 560-0043, 
Japan}
\author{Hiroshi Kohno} 
\affiliation{
Graduate School of Engineering Science, Osaka University, 
Toyonaka, Osaka 560-8531, Japan}

\date{\today}
\begin{abstract}
 A self-contained theory of the domain wall dynamics 
in ferromagnets under finite electric current is presented. 
 The current is shown to have two effects; 
one is  momentum transfer, which is proportional to the charge 
current and wall resistivity ($\rhow$), and the other is spin transfer, 
proportional to spin current. 
 For thick walls, as in metallic wires, the latter dominates 
and the threshold current for wall motion is 
determined by the hard-axis magnetic anisotropy, except for the case of very strong pinning. 
 For thin walls, as in nanocontacts and magnetic semiconductors, 
the momentum-transfer effect dominates, and 
the threshold current is proportional to $\Vz/\rhow$, $\Vz$ being the pinning potential.
\end{abstract}
\maketitle
Manipulation of magnetization and magnetic domain wall \cite{Hubert98} 
 by use of electric current is of special interest recently 
\cite{Slonczewski96,Tserkovnyak02,Klaeui03,Vernier03,Yamaguchi03},
from the viewpoint of application to spintronics, 
e.g., novel magnetic devices where the information is written electrically, 
and also as a basic physics in that it involves fascinating 
angular momentum dynamics.

 Current-driven motion of a domain wall was studied 
in a series of pioneering works by Berger 
\cite{Berger78,Berger84,Berger92}.
 In 1984, he argued that the electric current exerts a force 
on the domain wall via the exchange coupling \cite{Berger84}. 
 Later in 1992, he discussed that a spin-polarized current (spin current) 
exerts a torque on the wall magnetization, 
and studied the wall motion due to a pulsed spin-polarized 
current \cite{Berger92}.
 These theoretical works are, however, based on his deep physical insight, 
and seems to lack transparency as a self-contained theory. 
 Also, their phenomenological character makes the limit of applicability 
unclear. 
 In view of recent precise experiments \cite{Klaeui03,Vernier03,Yamaguchi03},
a general theory starting from a microscopic description is now needed.

 In this paper, we reformulate the problem of domain wall dynamics in the 
presence of electric current, and explore some new features 
such as current-induced depinning of the wall. 
 We start from a microscopic Hamiltonian with an exchange interaction 
between conduction electrons and spins of a domain wall \cite{Hydro}. 
 With a key observation that 
the position $X$ and polarization $\phiz$ of the wall 
are the proper collective coordinates \cite{Rajaraman82} 
to describe its dynamics, 
it follows straightforwardly that the electric current 
affects the wall motion in two different ways, 
in agreement with Berger's observation. 
 The first is as a force on $X$, or momentum transfer, 
due to the reflection of conduction electrons. 
 This effect is proportional to the charge current and wall resistance, 
and hence negligible except for very thin walls. 
 The other is as a spin torque (a force on $\phi_0$), arising when an 
electron passes through the wall. 
 Nowadays it is also called as spin transfer \cite{Slonczewski96}
between electrons and wall magnetization.  
 This effect is the dominant one for thick walls where the spin of the 
electron follows the magnetization adiabatically. 

 The motion of a domain wall under a steady current is studied in two 
limiting cases.
 In the adiabatic case, we show that even without a pinning force, there is 
a threshold spin current, $\jsc$, below which the wall does not move. 
 This threshold is proportional to $\Kp$, the hard-axis magnetic 
anisotropy. 
 Underlying this is that the angular momentum transferred from the electron 
can be carried 
by both translational motion ($X$) and polarization ($\phi_0$) of the wall, 
and the latter can completely absorb the spin transfer if the 
spin current is small, $ \js < \jsc$. 
 The pinning potential $V_0$ for the wall position ($X$) affects $\jsc$ 
only if it is very strong, $\Vz \gtrsim \Kp/\alpha$, where $\alpha$ 
is the damping parameter in the Landau-Lifshits-Gilbert equation.
 In most real systems with small $\alpha$, the threshold would thus be 
determined by $\Kp$. 
 Therefore, the critical current for the adiabatic wall
will be controllable by the sample shape and, in particular, 
by the thickness of the film, and does not suffer very much 
from pinning arising from sample irregularities. 
 This would be a great advantage in application.
 The wall velocity after depinning is found to be 
$\langle\dot{X}\rangle\propto \sqrt{(\js/\jsc)^2-1}$.

 In the case of thin wall, the wall is driven by the momentum transfer, which is proportional to the charge current $j$ 
and wall resistivity $\rhow$. 
 The critical current density in this case is given by 
$ \jc \propto \Vz/\rhow$.


 We consider a ferromagnet consisting of localized spins $\Sv$ and conduction 
electrons. 
 The spins are assumed to have an easy $z$-axis and a hard $y$-axis. 
 In the continuum approximation, the spin part is described by the Lagrangian 
\cite{Bouzidi90,BL96,TT96} 
\begin{eqnarray}
\lefteqn{ 
L_S = \int \frac{\dx}{a^3} 
    \bigg[\hbar S\dot\phi(\cos\theta -1)  -V_{\rm pin}[\theta] }  \\
&-& \frac{S^2}{2} 
    \Big\{ J \big( (\nabla\theta)^2 + \sin^2 \theta(\nabla\phi)^2 \big) 
   + \sin^2\theta \, ( K + K_\perp \sin^2\phi ) \Big\} \bigg], \nonumber
   \label{L00}
\end{eqnarray}
where $a$ is the lattice constant, and we put 
$\Sv (x) = S \, (\sin \theta \cos \phi, \sin \theta \sin \phi, \cos \theta)$, 
and $J$ represents the exchange coupling between localized spins. 
 The longitudinal (${K}$) and transverse (${K}_\perp$) anisotropy constants  incorporate the effect of demagnetizing field. 
 The constants $J$, ${K}$ and ${K}_\perp$ are all positive.
 The term $V_{\rm pin}$ represents pinning due to additional localized 
anisotropy energy. 
 The exchange interaction between localized spins and conduction electrons 
is given by
\begin{equation}
  H_{\rm int} 
 = - \frac{\Delta}{S} \int \dx \Sv({x}) \cdot (c^\dagger \sigmav c)_x
\end{equation}
where $\Delta$ and $c$ ($c^\dagger$) are the energy splitting and 
annihilation (creation) operator of conduction electrons, respectively, 
and $\sigmav$ is a Pauli-matrix vector. 
 The electron part is given by 
$H_{\rm el} = \sum_{\kv} \epsilon_{\kv} c^\dagger_\kv c_\kv$ 
with $\epsilon_{\kv} = \hbar^2 \kv^2 /2m$.
 In the absence of $V_{\rm pin}$ and $H_{\rm int}$, the spin part has a static 
domain wall of width $\lambda \equiv (J/K)^{1/2}$ 
as a classical solution. 
 We consider a wire with width smaller than $\lambda$, and treat  
the spin configuration as uniform in the $yz$-plane, 
perpendicular to the wire direction $x$.
 The solution centered at 
$x=X$ is given by $\theta = \theta_0(x-X)$, $\phi = 0$, where
$\cos\theta_0(x)$$ = \tanh({x}/{\lambda})$, and 
$\sin\theta_0(x) 
 =\left( \cosh({x}/{\lambda}) \right)^{-1}$.
 To describe the dynamics of the domain wall, it is crucial to observe that 
the weighted average of $\phi$, defined by
$ \phiz(t) \equiv \int\ (dx/2\lambda) \phi(x,t) \sin^2\theta_0(x-X(t)) $
plays the role of momentum conjugate to $X$, and hence must be treated as 
dynamical\cite{TT96}.
 Neglecting spin-wave excitations, we obtain the Lagrangian for $X(t)$ and 
$\phiz(t)$ as 
\begin{equation}
 L_S = - \frac{\hbar NS}{\lambda}X \dot{\phi}_0 
       - \frac{1}{2}K_\perp NS^2 \sin^2\phiz 
       - V_{\rm pin}(X),\label{L0} 
\end{equation}
where $V_{\rm pin}(X)$ is a pinning potential for $X$, and 
 $N = 2A \lambda/a^3$ 
is the number of spins in the wall. 
($\Area$ is the cross-sectional area.) 
 The equations of motion, derived from the Lagrangian, 
$L_S-H_{\rm int}$, are given by 
\begin{eqnarray}
     \frac{\hbar NS}{\lambda} 
 \left( \dot{\phiz} +\alpha\frac{\dot{X}}{\lambda} \right)
 &=& F_{\rm pin} + F_{\rm el}
\label{DWeq01} \\
  \frac{\hbar NS}{\lambda} (\dot{X}-\alpha \lambda\dot{\phiz}) 
 &=& \frac{NS^2 K_\perp}{2} \sin 2\phiz + T_{{\rm el}, z} ,
\label{DWeq02}
\end{eqnarray}
where $ F_{\rm pin} = - (\partial V_{\rm pin} / \partial X) $, 
\begin{equation}
  F_{\rm el} 
 \equiv - \frac{\Delta}{S} \int \dx \nabla_x \Sv_0(x-X) \cdot {\nv}(x),
\end{equation}
and 
\begin{equation}
 {\bm T}_{\rm el} 
 \equiv - \frac{\Delta}{S} \int \dx \, \Sv_0(x-X) \times {\nv}(x) .
\end{equation}
 Here $\Sv_0$ denotes $\Sv (x)$ with 
$\theta = \theta_0(x-X)$, $\phi = \phi_0$, 
and ${n}_\mu\equiv \langle c^\dagger \sigma_\mu c \rangle$ ($\mu=x,y,z$) 
is (twice) the spin density of conduction electrons. 
$F_{\rm el}$ represents a force acting on the wall, 
or momentum transfer, due to the electron flow, 
while ${\bm T}_{\rm el}$ is a spin torque, or spin transfer, which 
comes from the directional mismatch between wall magnetization 
$\Sv_0(x-X)$ and $\nv (x)$. 
 We have added a damping term ($\alpha$), which represents a standard 
damping torque (Gilbert damping), 
$\Tv_{\rm damp} = - \frac{\alpha}{S} \Sv \times \dot\Sv$\cite{Hubert98}.
 Note that the spin-transfer effect acts as a source to the wall velocity  
via $ v_{\rm el} \equiv (\lambda / \hbar N S) \, T_{{\rm el}, z}$.

 To estimate $F_{\rm el}$ and $v_{\rm el}$, we calculate spin polarization 
${\nv}(x)$ in the presence of a domain wall by use of a local gauge 
transformation in spin space 
\cite{TF94}, $c(x) = U(x) a(x)$, 
where $a(x)$ is the 2-component electron operator in the rotated frame, 
and $U(x) \equiv {\bm m}(x)\cdot\sigmav$ is an SU(2) matrix 
with 
$ {\bm m} (x) = \left(
  \sin\frac{\theta_0(x-X)}{2}\cos\phiz ,
  \sin\frac{\theta_0(x-X)}{2}\sin\phiz ,
  \cos\frac{\theta_0(x-X)}{2} \right)       $.
 The expectation value in the presence of electric
current is written in terms of the Keldysh Green function in the rotated 
frame. 
 For instance,
$ n_x(x) =  [(1-\cos\theta_0) \cos^2\phiz-1] \tiln_x  
          + (1-\cos\theta_0)\cos\phiz \sin\phiz \tiln_y 
          +  \sin\theta_0 \cos\phiz  \tiln_z $,
where
$ \tiln_\mu(x) \equiv 
-i{\rm Tr} (G_{xx}^<(t,t) \sigma_\mu) $, 
$G_{x\sigma,x'\sigma'}^< (t,t') \equiv 
i\langle a^\dagger_{x',\sigma'}(t') a_{x,\sigma}(t) \rangle$ 
being the lesser component of the Keldysh Green function \cite{Keldysh64}. 
 After a straightforward calculation, we obtain 
\begin{equation}
  F_{\rm el} = - \pi \hbar^2 \frac{\Delta}{L^2}
                 \sum_{\kv q\sigma} u_q^2 f_{\kv\sigma}
    \frac{(2k+q)_x}{2m} 
    \sigma \delta({\epsilon_{\kv+q,-\sigma}-\epsilon_{\kv\sigma}}),
\label{Feq}
\end{equation}
and 
\begin{equation}
  v_{\rm el} = \frac{\hbar \lambda^2}{NS} 
  \frac{ \Delta}{L^2}\sum_{\kv q \sigma} u_q^2 f_{\kv\sigma}\frac{(2k+q)_x}{2m}
  \frac{{\rm P} }{\epsilon_{\kv+q,-\sigma}-\epsilon_{\kv\sigma}},
\label{veq}
\end{equation}
to the lowest order in the interaction (with wall) 
$u_q \equiv 
 - \int dx e^{-iqx} \nabla_x \theta_0(x) = 
 \frac{\pi} {\cosh(\pi\lambda q/2)}$. 
 The distribution function $f_{\kv\sigma}$ 
specifies the current-carrying non-equilibrium state, 
and P means taking the principal value. 
 As is physically expected, $F_{\rm el}$ is proportional to the 
reflection probability of the electron, and hence to the wall 
resistivity, as well as the charge current. 
 In fact, adopting the linear-response form, 
$ f_{\kv\sigma} \simeq f^{0}(\epsilon_{\kv\sigma}) 
    - e\Ev\cdot\vv \tau (\partial f^0 / \partial \epsilon )$, 
as obtained from the Boltzmann equation 
($f^0$: Fermi distribution function, $\Ev$: electric field, 
$\vv = \hbar \kv/m$, 
$\tau$: transport relaxation time due to a single wall), 
we can write as $F_{\rm el} = enjR_{\rm w}$ in one dimension. 
 Here $n$ and $j$ are the electron density and current density, 
respectively, and 
$R_{\rm w} = \frac{h}{e^2} \frac{\pi^2}{8} \frac{\zeta^2}{1-\zeta^2}
             (u_+^2+u_-^2)$ 
is the wall resistance \cite{GT00}, 
with 
$\zeta \equiv (k_{F+}-k_{F-})/
              (k_{F+}+k_{F-})$ and 
$u_\pm \equiv u_{k_{F+}\pm k_{F-}}$.
 More generally, one can prove the relation \cite{KT04} 
\begin{equation}
  F_{\rm el} = e\Ne \rhow j = en R_{\rm w} I \Area ,
\label{FR}
\end{equation} 
using Kubo formula, 
where $\rhow \equiv R_{\rm w}\Area/L$ is the resistivity due to a wall 
\cite{rho}, $I\equiv jA$, and 
$\Ne\equiv nL\Area$ is the total electron number.  

 Equations (\ref{DWeq01}) and (\ref{DWeq02}), with (\ref{veq}) and (\ref{FR}) 
constitute a main framework of the present paper. 
 We next go on to studying them in the two limiting cases; 
adiabatic wall and adrupt wall.

 We first study the adiabatic limit, which is of interest for metallic nanowires, 
where $\lambda \gg k_{\rm F}^{-1}$. 
 In this limit, we take 
$ u_q^2 \rightarrow \frac{4\pi}{\lambda}\delta(q) $, 
and by noting 
$({\epsilon_{\kv+q,-\sigma}-\epsilon_{\kv\sigma}})_{q=0}=2\sigma\Delta\neq0$,
we immediately see from Eq.(\ref{Feq}) that $F_{\rm el}=0$, whereas 
\begin{equation}
v_{\rm el}
  = \frac{\lambda \hbar}{NS}\frac{1}{L} \sum_{\kv\sigma}\sigma 
    \frac{k_x}{m}f_{\kv\sigma} 
  = \frac{1}{2S}\frac{a^3}{e} \js 
  \label{velad}
\end{equation}
remains finite. 
 The spin transfer in this adiabatic limit is thus proportional 
to spin current flowing in the bulk (away from the wall), 
$ \js \equiv \frac{e\hbar}{mV}\sum_{\kv} k_x (f_{\kv+}-f_{\kv-})$ 
($V\equiv L\Area$ being the system volume).
 In reality, the spin current is controlled only by controlling charge 
current.
 In the linear-response regime, it is proportional to the charge current $j$ 
as 
$ \js = \eta \, j $, 
$\eta$ being a material constant. 
 This parameter can be written as 
$\eta = \sum_\alpha  (\sigma_+^\alpha - \sigma_-^\alpha) / 
        \sum_\alpha  (\sigma_+^\alpha + \sigma_-^\alpha) $
for a wire or bulk transport, and 
$\eta = \sum_\alpha  (N_+^\alpha - N_-^\alpha) / 
        \sum_\alpha  (N_+^\alpha + N_-^\alpha) $
for a nanocontact and a tunnel junction, where 
$\sigma_\pm^\alpha$ and $N_\pm^\alpha$ are band ($\alpha$) and 
spin ($\pm$) resolved electrical conductivity and density of states 
at the Fermi energy, respectively, of a homogeneous ferromagnet. 
 For bulk transport in transition metals (such as in wire), $\eta$ is 
expected to be small since $s$ electrons dominate the conduction. 
 For tunnel junctions, in contrast, it may be large 
($\sim 50$\% \cite{Monsma00}), 
since $d$ electron contribution will be dominant because of its large 
density of states.

 As seen from Eq.(\ref{Xdot}) below, 
the speed of the stream motion of the wall is roughly given by $v_{\rm el}$ 
 (except in the vicinity of the threshold $\jc$).  
 For a lattice constant $a\sim 1.5 \AA$ and current density  
$j=1.2\times 10^{12}$ [A/m$^2$] \cite{Yamaguchi03}, 
we have $a^3 j/e\sim 250$ [m/s].
 This speed is expected for strongly spin-polarized materials ($\eta \sim 1$)
such as half metals. 
 In transition metals, where the transport is dominated by $s$ electrons, 
$\eta$ would be a few orders of magnitudes smaller, say $\sim 0.01$, 
which may explain the observed value $\sim 3$ [m/s] \cite{Yamaguchi03}.  
 Current-driven wall velocity may thus be useful in determining polarization $\eta$, which is of fundamental importance in spintronics.

\begin{figure}[btp]
\includegraphics[scale=0.5]{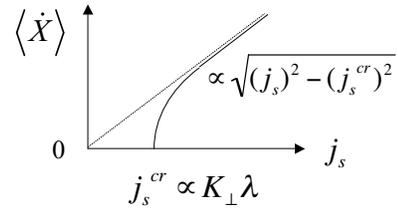}
\caption{ Time-averaged 
wall velocity as a function of spin current, $\js$, in the weak pinning case 
($\Vz\lesssim \Kp/\alpha$).
\label{FIGv-j}}
\end{figure}

 Let us study the wall motion in the absence of pinning, $F_{\rm pin}=0$, 
by solving the equations of motion, (\ref{DWeq01}) and (\ref{DWeq02})
in the adiabatic case ($F_{\rm el}=0$). 
 The solution with the initial condition $X = \phi_0 = 0$ at $t=0$ is 
obtained as 
\begin{eqnarray}
  \kappa \cot \left( \frac{\alpha}{\lambda}X \right) 
 &=&  \sqrt{1-\kappa^2} \, \coth (\gamma t) + 1 
\ \ \ (|\kappa| < 1) \\
 &=&  \sqrt{\kappa^2-1} \, \cot  (\omega t) + 1 
\ \ \ (|\kappa| > 1) 
\end{eqnarray}
where $\kappa \equiv 2\hbar v_{\rm el} / (SK_\perp\lambda)$, 
$\gamma = 
 \frac{\alpha}{1+\alpha^2} \frac{S K_\perp}{2\hbar} \sqrt{1-\kappa^2}$, 
and 
$\omega =
 \frac{\alpha}{1+\alpha^2} \frac{S K_\perp}{2\hbar} \sqrt{\kappa^2-1}$.
 For ${|v_{\rm el}|} < v^{\rm cr} \equiv {SK_\perp\lambda}/2\hbar$ 
(i.e., $|\kappa| < 1$), 
$\cot (\alpha X / \lambda )$ remains finite as $t \to \infty$, 
and the wall is not driven to a stream motion but just displaced by 
$\Delta X = \frac{\lambda}{2\alpha}\sin^{-1}{\kappa}$. 
 In this case, the transferred spin is absorbed by $\phiz$ and dissipated 
through $K_\perp$, as seen from Eq.(\ref{DWeq02}), 
and is not used for the translational motion of the wall ($\dot X$); 
the wall is apparently \lq\lq pinned'' by the transverse anisotropy.
 Thus even without pinning force, the current cannot drive the wall 
if the associated spin current is smaller than the critical value 
\begin{equation}
 \jsca =  \frac{e S^2}{a^3 \hbar} K_\perp \lambda . 
\end{equation}
 Above this threshold, $\js > \jsca $ ($|\kappa| > 1$), 
this process with $K_\perp$ cannot support the transferred spin 
and the wall begins a stream motion. 
 The wall velocity after \lq\lq depinning'' is an oscillating function 
of time around the average value  (Fig.\ref{FIGv-j}) 
\begin{equation}
 \langle \dot X \rangle 
 = \frac{1}{1+\alpha^2} \frac{1}{2S} \frac{a^3}{e} 
\sqrt{ \js ^2 - \big( \jsca \big)^2 },  
\label{Xdot}
\end{equation} 
which is similar to the Walker's solution for the field-driven case 
\cite{Hubert98}. 
 (The bracket $\langle \cdots \rangle$ means time average.)

 We now introduce a pinning potential, $V_{\rm pin}$,  
and study the \lq\lq true'' depinning of the wall by 
the spin-transfer effect in the adiabatic limit. 
 Since spin transfer acts as a force on $\phiz$, 
the depinning can be better formulated in terms of $\phiz$.
 We consider a quadratic pinning potential with a range $\xi$; 
\begin{equation}
 V_{\rm pin} = \frac{N\Vz}{\xi^2} (X^2 -\xi^2)\theta(\xi-|X|),
\label{Vpin}
\end{equation}
where 
$\theta (x)$ is the Heaviside step function. 
 Then the equation for $\phiz$ reads 
\begin{equation}
(1+\alpha^2)\ddot{\phiz}
  = -\alpha \dot{\phiz} (\nu + \mu \cos2\phiz)
    - \nu \left( \frac{\mu}{2}\sin2\phiz 
    + \frac{v_{\rm el}}{\lambda}\right) ,
\end{equation}
where $\mu \equiv S \Kp/\hbar $ and
$\nu \equiv 2\Vz\lambda^2/\xi^2 \hbar S$. 
 This equation describes the motion of a classical particle 
in a tilted washboard potential $\tilde V$ with (modified) friction. 
 For 
$v_{\rm el} > v^{\rm cr} (= \mu \lambda / 2)$, 
local minima disappear in $\tilde V$ and $\phiz$ is then \lq\lq depinned''.
 Then Eq.(\ref{DWeq02}) indicates that $\phiz$ starts to drift with average 
velocity 
$\langle \dot\phiz \rangle = - v_{\rm el} / (\alpha\lambda)$, 
with oscillating components superposed. 
 The time development of $X(t)$ is then obtained as
$ X = (v_{\rm el} / \nu \alpha) (1 - e^{- \nu t/\alpha})$ 
for $|X| \leq \xi$.
 The depinning of the wall occurs when $X(t)$ reaches $\xi$. 
 Thus, the critical spin current $\jsc$ for depinning is given by 
$\jsca$ defined above if the pinning is weak 
($\Vz \lesssim \Kp/\alpha$), 
while it is given by
\begin{equation}
 \jscb \equiv  \frac{4e}{a^3\hbar} \alpha\Vz\lambda,
\end{equation}
if the pinning is strong, $\Vz \gtrsim \Kp/\alpha$.
 Since $\alpha$ is usually believed to be small \cite{Berger92}, we expect
that the critical current is mostly determined by $\Kp$.
 This seems to be consistent with the observations that the critical 
current is larger for a thinner film 
\cite{Yamaguchi03,Berger92}.
 It would be interesting to carry out measurements on a wire with small $\Kp$, 
such as that with a round cross section. 

Let us go on to the opposite limit of abrupt wall, $\lambda\rightarrow 0$.
As seen from Eq.(\ref{veq}), the spin-transfer effect vanishes. 
 The pinning-depinning transition is thus determined by the competition 
between $F_{\rm el} = e \Ne \rhow j$ and $F_{\rm pin}$. 
 It occurs when $F_{\rm el} = N \Vz/\xi$, giving the critical current density 
\begin{equation}
 \jc = \frac{N\Vz}{\xi e\Ne\rhow} 
     = \frac{2\Vz\lambda}{ena^3\xi R_{\rm w}A}.
\end{equation}
 The average wall velocity after depinning is obtained as
$ \langle \dot X \rangle 
 = \frac{\lambda^2 \Ne e}{\hbar\alpha NS}\rhow j 
 = (ena^3/2\hbar S)(I R_{\rm w})/\alpha$.
 This velocity vanishes in the limit, $\lambda \to 0$, 
due to the divergence of the wall mass, 
$M_{\rm w} = \hbar^2 N /(\Kp \lambda^2)$.

 For metallic nanocontacts, where
$\xi\sim \lambda\sim a$\cite{Garcia99} and $na^3\sim1$, 
experiments indicate that the wall resistance can be of the order of 
$h/e^2 = 26$ k$\Omega$ \cite{Garcia99}.
 Thus 
$ \jc \sim 
(5\times 10^{10}\times B_c$[T])[A/m$^2$], 
where $ B_c=\Vz\lambda/\mu_{\rm B} \xi S$ is the depinning field ($ \mu_{\rm B}$ is Bohr magneton).
$ B_c \sim 10^{-3}$ [T] 
(like in Ref.\cite{Garcia99}) corresponds to
$ \jc \sim 5\times 10^7$ [A/m$^2$].

 In conclusion, we have developed a theory of domain wall dynamics 
including the effect of electric current. 
 The current is shown to have two effects; 
spin transfer and momentum transfer, as pointed out by Berger. 
 For an adiabatic (thick) wall, where the spin-transfer effect due to 
spin current is dominant, there is a threshold spin current 
$\jsc \sim (e\lambda / a^3){\rm max} \{ K_\perp, \alpha V_0 \}$
below which the wall cannot be driven.  
 This threshold is finite even in the absence of pinning potential. 
The wall motion is hence 
not affected by the uncontrollable pinning arising from sample roughness
if weak-pinning ($\Vz \lesssim \Kp/\alpha$). 
In turn, wall motion would be easily controlled by the sample shape, 
through the demagnetization field and thus $\Kp$. 
 The wall velocity after depinning is obtained as 
$ \langle \dot X \rangle \propto \sqrt{(\js)^2-(\jsc )^2}$. 
 In contrast, an abrupt (thin) wall is driven by the momentum-transfer 
effect due to charge current, i.e., by reflecting electrons. 
 In this case, the depinning current is given in terms of wall resistivity 
$\rhow$ as $ \jc \propto \Vz/\rhow $.

 The two limiting cases considered above are both realistic.
 Most metallic wires fabricated by lithography are in the adiabatic limit, 
as is obvious from very small value of wall resistivity \cite{Kent01}.
 In contrast, very thin wall is expected to be formed in metallic magnetic 
nanocontacts with a large magnetoresistance (called BMR) \cite{Garcia99}.
 A system of recent interest is magnetic semiconductors \cite{Ohno98}, 
where the Fermi wavelength is much longer than in metallic systems. 
 As suggested by large magnetoresistance observed recently \cite{Ruester03}, 
magnetic semiconductors would be suitable for precise measurement in the thin 
wall limit. 


 The authors are grateful to T. Ono for motivating us by showing the 
experimental data prior to publication. 
 We also thank J. Shibata and A. Yamaguchi for valuable discussions.
 G.T. is grateful to Ministry of Education, Culture, Sports, Science and
Technology, Japan and The Mitsubishi Foundation for financial support.



\begin{thebibliography}{99}
\bibitem{Hydro}
We neglect the effect of hydromagnetic drag and classical Oersted field, 
which are small in thin wires\protect\cite{Klaeui03}.
\bibitem{rho}
$\rho_{\rm w}$ is proportional to the wall density, i.e., $1/L$, and so 
$N_e \rho_{\rm w}$ is independent of $L$.
\end{thebibliography}
\end{document}